\documentclass[11pt]{article}
\usepackage{graphicx}
\usepackage{amsmath,amssymb}
\newcommand{\comment}[1]{{}}

\title{Topological excitations around the vacuum of quantum gravity I: The symmetries of the vacuum}

\author{Artem Starodubtsev\thanks{email: astarodu@astro.uwaterloo.ca} \\
 \\ \centerline{\footnotesize \it Department of Physics, University of Waterloo} \\ \centerline{\footnotesize \it 200
University ave W, Waterloo, ON, Canada N2L 3G1} \\
\centerline{\footnotesize \it and} \\
\centerline{\footnotesize \it Perimeter Institute for
Theoretical Physics} \\
\centerline{\footnotesize \it 35 King st N, Waterloo, ON, Canada
N2J 2W9 }}

\date{}

\begin{document}

\maketitle

\begin{abstract}
This is the first paper in a series in which an attempt is made to
formulate a perturbation theory around the the Chern-Simons state
of quantum gravity discovered by Kodama. It is based on an
extension of the theory of 't Hooft Deser and Jackew  describing
point particles in 3D gravity to four spacetime dimensions.
General covariance now requires the basic excitations to be
extended in one spatial dimension rather than pointlike. As a
consequence the symmetry of the Kodama state, which is the
(anti)deSitter symmetry, is realized 'holographically' on a
timelike boundary, the generators of this symmetry being related
to quasilocal energy-momenta. As GR induces a Chern-Simons theory
on the boundary, the deSitter symmetry of the vacuum must be
q-deformed with the deformation parameter related to the
cosmological constant.  It is proposed to introduce excitations
around this vacuum by putting punctures on the boundary to each of
which is associate a vector in some representation of deSitter
group projected to the boundary. By equations of motion those
punctures must be connected by continuous lines of fluxes of an
SO(3,1) gauge field. It is also argued that quasilocal masses and
spins of these excitations must satisfy a relation of Regge type,
which may point on a possible relation between non-perturbative
quantum gravity and string theory.

\end{abstract}

\section{Introduction}
The main problem of loop quantum gravity (see \cite{reviewrt} for
review) is the problem of the classical limit. Soon after the
hamiltonian constraint of quantum gravity  was defined on spin
network states \cite{Thiemann96}, and a wide class of solutions to
it was found, it has become apparent that none of those solutions
represent continuous spacetime geometry in the classical limit.
The problem manifests itself in the fact that the hamiltonian
constraint acts on distinct vertices of a spin network
independently allowing for no possibility to send a signal between
them \cite{Smolin96}. Also, the algebra of the hamiltonian
constraints appears to be abelian and therefore not isomorphic to
the classical constraints algebra \cite{LewandowskiMarolf}.

There are two possible strategies to address this problem. First
one can try to approximate states of the type that have been
proved to have a good classical limit in ordinary QFT by
complicated combinations of loop or spin network solutions. This
includes the coherent states program \cite{Thiemann99} and
approximating states of the Fock space \cite{Varadarajan2000}.
Second one can extend the range of our search to include other
states which are not of loop or spin network type.

An exact state of quantum gravity which has a good classical limit
does exist and this is the Chern-Simons state discovered by Kodama
back in 1990 \cite{Kodama90}.
\begin{equation}
\Psi_K = e^{\frac{3}{2\Lambda G \hbar}S_{CS}(A)},
\end{equation}
where $S_{CS}(A)$ is the Chern-Simons action for the Ashtekar
connection \cite{Ashtekar}, and $\Lambda$ is the cosmological
constant.
 Although we don't have a strict
definition of the physical space of states the Kodama state
belongs to and, therefore, we cannot calculate the expectation
values of physical observables to show that they are peaked around
a classical solution, there is some  indirect evidence that the
Kodama state has a good classical limit. On each spacial slice it
satisfies a self-dual set of initial conditions (electric field is
proportional to magnetic). These initial conditions are preserved
by Einstein's equations and result in the deSitter or
anti-deSitter spacetime depending on the sign of the cosmological
constant. This means that (anti)deSitter spacetime must be the
classical limit of the Kodama state. For more details about the
Kodama state, its relation to loop states and an outline of the
program of how to solve the problem of classical limit in loop
quantum gravity by using it see \cite{Smolin02}.

The problem with the Kodama state is that there is only one state
of this kind known. The question is then: Can we define states
which can be considered as perturbations around the Kodama state,
the later playing the role of the vacuum? More precisely: Given
that the classical limit of the Kodama state is the deSitter
spacetime can we define states the classical limit of which would
be gravitons propagating against the deSitter background?

This is not an easy question to answer because to define these
excitations explicitly we need some spacetime structure to build
them on. But we don't know a priori what quantum spacetime is. One
can setup a semiclassical framework \cite{Smolin02} in which
excitations around the Kodama state are introduced  via expanding
around  classical deSitter spacetime. This has already led to some
nontrivial predictions such as modified energy-momentum relations
which cannot be obtained from studying purely classical spacetime.
The next step would be to develop a formalism that does not rely
on any classical spacetime structure at all. One may hope that
such completely background independent consideration may reveal
more information on what quantum spacetime is. This paper is about
a particular way of approaching this.

In section 2 it is suggested that the structure of a TQFT which
the Kodama state is the only solution of can be used to define
excitations. This can be done by analogy with 't Hooft, Deser, and
Jackew's formulation of 2+1 gravity with point particles
\cite{HooftDJ84}. The new features arising in 3+1 dimensions are
described.

Among those features is a 'holographic' realization of the basic
symmetries and observables of the theory. This requires a specific
formulation of GR in the presence of a (timelike) boundary. This
is done in section 3.

In section 4 the boundary theory is studied in more details. The
algebra of boundary observables is derived and the generators of
that algebra are related to quasilocal energy momentum and spin.
The formalism fixes the coupling constant of the boundary
Chern-Simons theory which leads to a q-deformation of the symmetry
of the ground state.

In section 5 all the constraint of general relativity are recast
in a form of a local conservation law. This form of constraints
suggests how the boundary observables should be continued into the
bulk.

In the absence of a strict regularization scheme some generic
properties of the boundary observables are studied in section 6.
In particular it is shown that the Kodama state is the ground
state.

In section 7 a general way of introducing excitations around the
Kodama state is discussed. In some simple cases the spectra of the
boundary observables can be evaluated. In particular it is argued
that quasilocal masses and spins must satisfy a relation of the
Regge type.

\section{Point particles in 2+1 gravity: how to extend to 3+1 dimensions}
It is remarkable that in 2+1 gravity there is a way to define
excitations around the vacuum which does not refer to any
structure of classical spacetime. This is a specific way of
describing point particles moving in 2+1 dimensional spacetime.
Below for definiteness we consider the case of positive
cosmological constant.

The action of 2+1 gravity is
\begin{equation}
S=\int (e\wedge F(A) + \Lambda e\wedge e\wedge e), \label{gr2+1}
\end{equation}
where the connection $A$ and the metric $e$ take the values on 2+1
dimensional $\gamma$-matrices and $F(A)$ is the curvature of the
connection $A$. The action (\ref{gr2+1}) can also be considered as
the Chern-Simons action for $SO(3,1)$ group. The equations of
motion of (\ref{gr2+1})
\begin{equation}
F(A)+\Lambda e\wedge e =0, \label{gr2+1c1}
\end{equation}
and
\begin{equation}
\nabla \wedge e=0 \label{gr2+1c2}
\end{equation}
simply mean that the curvature is constant everywhere and is
determined by the cosmological constant. Put in the Chern-Simons
language this means that $SO(3,1)$-curvature is zero. Therefore
there is no intrinsic dynamics in the theory.

Dynamics can be introduced by adding matter to the action. The
simplest example of matter is point particles having a certain
mass. In 2+1 dimensions point particles do not create a Newtonian
potential around them -- the  $SO(3,1)$ curvature is non-zero only
at  the points of the location of the particles. Classically these
particles can be represented as conical singularities with the
deficit angles proportional to their masses moving along certain
trajectories in the deSitter space.

Quantum mechanically, on the other hand, there are no definite
trajectories and the particle is described by a vector in a
certain representation of the $SO(3,1)$ group. Therefore from the
Chern-Simons point of view they can be thought of as point charges
of the $SO(3,1)$ gauge group, or topological punctures. It is
important to notice that $SO(3,1)$-symmetry is now understood as
the gauge symmetry, the symmetry generated by the constraints
(\ref{gr2+1c1},\ref{gr2+1c2}), and not as the global symmetry of a
classical background manifold.

This is an example of how the same object can be given two
equivalent descriptions, one of which is background dependent and
the other is background independent. On one side we have (quantum)
particles propagating against a classical deSitter background. On
the other side we have a topological field theory on a punctured
manifold with all the physical information encoded in the vectors
associated to the punctures and the positions of the punctures
being irrelevant.

It is tempting to try to extend the above picture to 3+1
dimensions which would give rise to excitations around the Kodama
state. The Kodama state is a solution not only of quantum general
relativity but also of a four dimensional topological field theory
called BF-theory defined by the following action
\begin{equation}
S=\int B\wedge F(A)+ \frac{\Lambda}{2} B\wedge B, \label{bfth}
\end{equation}
where $B$ is a 2-form field. The constraints of (\ref{bfth})
\begin{equation}
F(A)+\Lambda B =0 \label{bfthc}
\end{equation}
 have a unique solution, which is the Kodama state. So
 perturbations around the Kodama state can be considered as
 addition of external sources of curvature to the BF theory, in
 the same way as matter can be added to 2+1 gravity. The
 difference is that these sources  can still be intrinsic for 3+1 general
 relativity, i.e. still satisfy the vacuum Einstein equations.
 This is why pure 3+1 gravity has nontrivial intrinsic dynamics.

There are some other differences between the dynamics of
topological matter in 2+1 and in 3+1 dimensions. Let us first ask
why the natural   extrinsic sources in 2+1 gravity are point
particles or 0+1 dimensional objects. What makes the point
particles preferred over singular sources of other dimensions? The
reason is that this is the only possible objects that can be
defined in a general-covariant way. Point particles are sources of
curvature and the curvature is a 2-form. For the source to be
singular it must contain a $\delta$-function and the
$\delta$-function is a density. Density in turn results from the
duality operation acting on forms, and so the  volume form of the
source must be dual to the curvature form. Therefore the objects
which source the curvature must be of codimension 2, which are
point particles or punctures. These arguments are similar to those
according to which the electric charge in 11-dimensional M-theory
must be carried by M2-branes and magnetic charge by M5-branes. The
difference is that what is distributional now is not a divergence
of a curvature but the curvature itself.

The above logic extends directly to 3+1 dimensions but now objects
of codimension 2 are 1+1 dimensional objects. Classically the
natural candidate for such  objects would be  cosmic strings,
which are conical singularities extended in one spacial dimension.
Being stretched in flat space along $z$-axis they can be described
by the following metric
\begin{equation}
ds^2=-dt^2+dz^2 +dr^2 +(1-\alpha)^2 r^2 d\phi^2,
\end{equation}
where $\alpha$ is the deficit angle. This metric solves the
Einstein's equations with the following stress-energy tensor
\begin{equation}
T^{\mu \nu}=diag(\alpha,0,0,-\alpha)\delta(r), \label{tcs}
\end{equation}
which in particular means that it solves all the constraints
except the Hamiltonian. This means that general configurations of
cosmic strings are not purely gravitational objects as the do not
solve the vacuum Einstein's equations. Yet there may be some
specific configurations of them which solve the vacuum Einstein's
equations at least approximately in a weak coupling limit. The
example is the graviton mode of a string.

In this paper we will focus mostly on quantum mechanical
background independent description of the excitations. The natural
candidates for these are Wilson loops
\begin{equation}
W(\Gamma)=P e^{\int A_a dx^a}.
\end{equation}
They were shown to solve the gauge and the diffeomorphism
constraints and also the Hamiltonian constraints for a certain
factor ordering \cite{JacobsonSmolin88}, \cite{RovelliSmolin88}.
However the Kodama state is a solution to the Hamiltonian
constraint for a different factor ordering. Therefore, as
excitations around the Kodama state loops do not solve all the
constraints of GR. But, again,  there may exist specific
configurations of loops which solve all of them. They can be
formally defined via expansion of the projector on the physical
states or in spin foam models \cite{Rovelli98}.

Finally, the most profound difference between 2+1 and 3+1 TQFT is
that in the constraint algebra. The constraints of 2+1 gravity
(\ref{gr2+1c1},\ref{gr2+1c2})form an $SO(3,1)$-algebra, so the
symmetry of the only classical solution is present in the
canonical structure of the theory. On the other hand the algebra
of the constraints (\ref{bfthc}) of 3+1 dimensional TQFT is
abelian, while the symmetry of the only classical solution of the
theory which is the deSitter spacetime is $SO(4,1)$. So the
symmetry of the ''vacuum'' solution is not represented in the
constraint algebra. Behind this difference is the
''holographical'' nature of 3+1 dimensional gravity. This is a
consequence of the fact that the basic excitations of the theory
are non-local.

Let us expand the last point in a bit more detail. If the
constraints form the symmetry algebra of the vacuum classical
solution then they can be understood as operators of energy
momentum and angular momentum of the theory. Therefore in 2+1
gravity, as it is described here, the operators of the above
quantities can be introduced locally. Indeed, energy, momentum,
and spin can be assigned to each point particle, and because there
is no Newtonian potential, and no gravitational waves in the
theory, there is nothing like gravitational energy which generally
cannot be defined locally. In 3+1 dimensions on the other hand
excitations  which do not produce Newtonian potential and can in
principle represent the gravitational radiation themselves are
 strings. But these objects cannot be localized on spacelike
3-slice and do not allow us to define energy, momentum, and
angular momentum locally. This is why the symmetry of the vacuum
solution cannot be realized as the algebra of bulk constraints of
the theory. It can however be realized in a ''holographic'' sense.
If we pick out a closed 2-surface in the spacial slice and impose
a certain boundary conditions on it, then strings can be localized
at this surface at the intersection points. On the surface can be
defined a field theory, the constraints of which form  (a
subalgebra of) the deSitter algebra.

All the above fits very well with the well known fact that energy,
momentum and angular momentum cannot be defined locally in general
relativity. They however can be defined quasilocally on a surface
on which some boundary conditions are satisfied. This is done in a
form most suitable for our purposes in the next section.

\section{Action principle for GR with a boundary: the choice of
the boundary action}

As  was mentioned in the previous section, the only context in
which the excitations can be studied from the point of view of the
symmetry of the vacuum is when spacetime has a boundary. The
boundary doesn't have to surround the whole spacetime, it may be
finite, but a certain kind of condition on the boundary must be
satisfied.

In general the variation of an action for general relativity in a
bounded region contains a surface term which must vanish for the
variational principle to be well defined. This can be achieved by
imposing some boundary conditions plus adding a surface term to
the action.

There are many possible choices of the surface term in the action.
 Different boundary conditions require different surface terms.
But they do not determine them completely. The surface term also
depends on the choice of canonical variables. When we perform a
canonical transformation in the theory, this is equivalent to the
addition of a total derivative term to the Lagrangian. This term
doesn't change the equations of motion in the bulk, but it may
have a nontrivial effect in the boundary. This is another piece of
ambiguity in the choice of the surface term.

Here we will follow the proposal of Smolin
\cite{Smolin95,Smolin98} relating quantum gravity in the bulk
region with topological quantum field theory on the boundary. The
self-dual boundary conditions considered there were shown to be
satisfied on black hole horizons \cite{Krasnov96} and therefore
were extremely useful for studying black hole mechanics. However
they are not very suitable for our purposes. They do not allow us
to define energy and momentum and as a consequence  the boundary
theory does not completely capture the symmetry of the vacuum.

Therefore we choose another set of boundary conditions. To have a
sensible definition of energy and momentum we need to fix metric
on the boundary. We must also choose what canonical variables to
use and this is motivated by our interest in the Kodama state.
First, we must choose the Ashtekar variables in their original
form, with the Immirzi parameter equal to $i$, as only in this
variables the constraints take a simple form which the Kodama
state is a solution of. Second, it is well known that General
relativity can be obtained from breaking $SO(4,1)$-symmetry in
topological field theory down to $SO(3,1)$ \cite{mmons}. The
resulting form of GR action has definite implications as to what
the boundary theory should be. Given that $SO(4,1)$ is the
symmetry of the Kodama state it is natural to choose this form of
the boundary action.

Here it is worth reviewing how a breaking of $SO(4,1)$ symmetry in
TQFT leads to General Relativity, and what kind of action it
results in. For convenience, we will use spinorial notations. Our
starting point will be a topological field theory for $Sp(4)$
group which is locally isomorphic to $SO(4,1)$. The action depends
on $Sp(4)$-connection $A_\alpha^\beta$, where
$\alpha,\beta=0,..3$.
\begin{equation}
S=\frac{1}{8 \pi G\Lambda}\int Tr F^\alpha_\beta P^\beta_\gamma
\wedge F^\gamma_\delta P^\delta_\alpha. \label{action0}
\end{equation}
Here $F^\alpha_\beta=dA^\alpha_\beta+A^\alpha_\gamma \wedge
A^\gamma_\beta$ is the $Sp(4)$-curvature 2-form and $
P^\alpha_\beta$ is a fixed symmetry breaking 0-form matrix.

 $Sp(4)$ quantities can be decomposed by using $SL(2,C)$
 indices notation $A=0,1$,
$A'=0,1$. The $Sp(4)$ connection can be represented as
\begin{eqnarray}
A_\alpha^\beta= \left(
\begin{array}{cc}
A_{A}^B & \frac{1}{l}e_{A}^{B'} \\
\frac{1}{l}e_{A'}^B & A_{A'}^{B'}
\end{array} \right), \label{adecomp}
\end{eqnarray}
where $A_{A}^B$ and $A_{A'}^{B'}$ are left-handed and right-handed
$SL(2,C)$-connections respectively, $e_{A'}^B$ is a tetrad and $l$
is a fixed parameter of the dimension of length. Similarly one can
decompose $Sp(4)$-curvature
\begin{eqnarray}
F_\alpha^\beta= \left(
\begin{array}{cc}
F_{A}^B & F_{A}^{B'} \\
F_{A'}^B & F_{A'}^{B'}
\end{array} \right). \label{fdecomp}
\end{eqnarray}
Here
\begin{equation}
F_{A}^B=f_{A}^B+\frac{1}{l^2}e_{A}^{A'}\wedge e_{A'}^{B},
\label{fspfour}
\end{equation}
where $f_{A}^B$ is $SU(2)_L$-curvature of the connection $A_A^B$,
and
\begin{equation}
F_{A'}^B=\nabla \wedge e_{A'}^B
\end{equation}
is the torsion.

To get an action the canonical form of which is Ashtekar's one has
to restrict it to purely self-dual $SL(2,C)$-connection, which
means that we should choose $ P^\alpha_\beta$ in (\ref{action0})
to be
\begin{eqnarray}
P_\alpha^\beta= \left(
\begin{array}{cc}
\delta_{A}^B & 0 \\
0 & 0
\end{array} \right). \label{pdecomp}
\end{eqnarray}
The resulting bulk action is then
\begin{equation}
S_{Bulk}=\frac{1}{8 \pi G\Lambda}\int (f_{A}^B+\Lambda
e_{A}^{A'}\wedge e_{A'}^{B}) \wedge (f_{B}^A+\Lambda
e_{B}^{A'}\wedge e_{A'}^{A}), \label{action1}
\end{equation}
where $\Lambda=\frac{1}{l^2}$ is the cosmological constant. This
will be the basic bulk action for the rest of the paper.

As we mentioned before in order to be able to define energy and
momentum we choose to fix the metric on the boundary
\begin{equation}
\delta e_{A'}^{A} \Big\vert_S=0. \label{bcond}
\end{equation}
We can now calculate the variation of the action (\ref{action1})
subject to this condition.
\begin{eqnarray}
\delta S_{Bulk} &=& \frac{1}{4 \pi G\Lambda}\int (f_{A}^B+\Lambda
e_{A}^{A'}\wedge e_{A'}^{B})\wedge e_{B}^{B'} \wedge \delta
e^{A}_{B'} - \frac{1}{4 \pi G\Lambda}\int \nabla \wedge
(f_{A}^B+\Lambda e_{A}^{A'}\wedge e_{A'}^{B}) \wedge \delta A^A_B
\nonumber  \\
&+& \frac{1}{4 \pi G\Lambda} \int_S (f_{A}^B+\Lambda
e_{A}^{A'}\wedge e_{A'}^{B})\wedge \delta A^A_B. \label{abvar}
\end{eqnarray}
For the variation principle to be well defined the boundary term
in (\ref{abvar}) has to vanish which can be  achieved by adding a
boundary term to the action (\ref{action1}):
\begin{eqnarray}
S=S_{Bulk}+S_S, \label{action2}
\end{eqnarray}
where
\begin{eqnarray}
S_S=\frac{1}{6\pi G\Lambda}S_{CS}[A]+\frac{1}{4 \pi G}\int_S
e_{A}^{A'}\wedge A^A_B \wedge e_{A'}^{B}+ S[e], \label{actions}
\end{eqnarray}
where $S_{CS}[A]$ is Chern-Simons action of the connection $A$ and
$S[e]$ is a surface action which depends purely of metric. It can
be checked that the the variation of the action (\ref{action2})
subject to the condition (\ref{bcond}) has no surface term for
arbitrary choice of $S[e]$ in (\ref{actions}). The later can be
fixed by the requirement that the total action (\ref{action2}) be
covariant, i.e. that the gauge and diffeomorphism invariance at
the boundary be broken by the boundary conditions and not by the
action itself. However diffeomorphism and gauge symmetries are
partially broken due to the presence of the boundary. We can keep
only invariance with respect to diffeomorphisms tangent to the
boundary and with respect to rotations in the tangent space of the
boundary. To make it explicit let us introduce an arbitrary unit
vector field on the boundary $s^\mu$ and its spinorial
representation $s^A_{A'}=s^\mu e^A_{\mu A'}$,
$s^A_{A'}s^{A'}_{B}=\delta^A_B$. By using it one can parametrize
the self-dual part of the tetrad  $e^A_{A'}\wedge e^{A'}_{B}$ by a
traceless triad
$\sigma^A_B=e^A_{A'}s^{A'}_B-1/2e^C_{A'}s^{A'}_C\delta^A_B$ as
\begin{equation}
e^A_{A'}\wedge e^{A'}_{B}=\sigma^A_C \wedge \sigma^C_B+s\wedge
\sigma^A_B.
\end{equation}
 The second term in the r.h.s. of (\ref{actions}) can then be
rewritten in the form
\begin{equation}
\int_S e_{A}^{A'}\wedge A^A_B \wedge e_{A'}^{B}= \int_S
\sigma_{A}^{B}\wedge A^A_C \wedge \sigma_{B}^{C}+\int_S s\wedge
A^A_B \wedge \sigma_{A}^{B}.\label{sti}
\end{equation}
The second term in the r.h.s of (\ref{sti}) does not admit a
covariant extension and has to be removed. This can be done by
choosing the vector field $s$ on the boundary to be the unit
normal to this boundary, which makes the above term disappear
automatically. This means that the triad $\sigma^A_B$ is chosen to
be the projection of the tetrad $e^A_{A'}$ on the surface $S$. The
remaining term in the r.h.s of (\ref{sti})  by a specific choice
of $S[e]$ can be completed to the term with a covariant derivative
of $\sigma$. The covariant form of the boundary action is thus
\begin{eqnarray}
S_S=\frac{1}{6 \pi G \Lambda}S_{CS}[A]+\frac{1}{4 \pi G}\int_S
\sigma_{A}^{B}\wedge \nabla \wedge \sigma_{B}^{A} \label{actionsc}
\end{eqnarray}

In \cite{Smolin98} two possible sets of boundary conditions dual
to each other were studied. One could either fix connection on the
boundary $\delta A_A^B \bigg\vert_S=0$ or choose self-dual
boundary conditions $F_A^B-\Lambda e_A^{A'}\wedge e_{A'}^B$ and
leave the connection loose. Similarly, instead of fixing metric on
the boundary the action principle (\ref{action2}) by imposing the
following set of conditions
\begin{equation}
\nabla\wedge \sigma^A_B \bigg\vert_S =0 \label{e0}
\end{equation}
can be made consistent. As in the case of free varying laps and
shift functions we cannot define energy and momentum, the
condition (\ref{e0}) must imply that energy and momentum is zero.
In the next section we will see that this is indeed the case.

\section{Quasilocal quantities and the algebra of boundary observables}
In this section we will study the boundary theory defined by the
action (\ref{actionsc}) in more details and relate its observables
with quasilocal energy, momenta, and angular momenta of the bulk
theory.

A theory of the form (\ref{actionsc}) was considered by Witten
\cite{witten89} along with the action of 2+1 gravity. It can be
rewritten as a Chern -Simons action
\begin{equation}
S_S=\frac{1}{6\pi G \Lambda} S_{CS}(a)
\end{equation}
for SO(3,1)-connection
\begin{equation}
a=A_i {\cal J}^i + \sqrt{\Lambda}\sigma_i {\cal P}^i,
\label{so31c}
\end{equation}
where
\begin{equation}
[{\cal J}^i,{\cal J}^j]= \epsilon^{ijk} {\cal J}_k, \ [{\cal
J}^i,{\cal P}^j]= \epsilon^{ijk} {\cal P}_k, \ [{\cal P}^i,{\cal
P}^j]= \epsilon^{ijk} {\cal J}_k
\end{equation}
are the generators of the SO(3,1) group. This means that the
constraints of the theory (\ref{actionsc}) form an SO(3,1) algebra
with respect to the boundary simplectic form. In this the theory
is similar to 2+1 dimensional gravity. Also its constraints have
the same form
\begin{eqnarray}
C^A_B&=&\epsilon^{\alpha \beta}(F_{\alpha \beta
B}^{A}+\Lambda \sigma_{\alpha C}^{A}\sigma_{\beta B}^{C}) \nonumber \\
H^A_B&=&\epsilon^{\alpha \beta}\nabla_{\alpha}\sigma_{\beta
B}^{A}, \label{bconst}
\end{eqnarray}
were indices $\alpha,\beta=1,2$ are two-dimensional spacial
manifold indices and $\epsilon^{\alpha \beta}$ is completely
antisymmetric tensor. It differs however form (2+1) gravity by the
fact that the gauge and the diffeomorphism constraints have traded
places. Also different are the canonical commutation relations
between basic variables. Now they are
\begin{eqnarray}
\{A^A_{\alpha B},A_{\beta D}^{C} \} &=&3 \pi G \Lambda
\epsilon_{\alpha \beta
}(\delta^A_D\delta^C_B+\epsilon^{AC}\epsilon_{BD}) \nonumber
\\
\{\sigma^A_{\alpha B},\sigma_{\beta D}^{C} \} &=&2\pi G
\epsilon_{\alpha \beta
}(\delta^A_D\delta^C_B+\epsilon^{AC}\epsilon_{BD}) \nonumber
\\
\{A^A_{\alpha B},\sigma_{\beta D}^{C} \} &=&0. \label{bcomrel}
\end{eqnarray}

The fact that the boundary theory is a Chern-Simons theory for
SO(3,1) group with the coupling constant $\kappa=\frac{6 \pi
}{G\Lambda\hbar}$ means that the symmetry group of the vacuum is
now q-deformed with $q=\exp(\frac{1}{\kappa+2})$. This means that
particles inserted in punctures of the boundary theory will
"propagate" in q-deformed spacetime.

In the rest of this section we will relate the constraints
(\ref{bconst}) with quasilocal observables of the bulk theory.
This relation will involve projection of spinors on surfaces which
may be spacelike or timelike. For this some useful notations are
introduced below.

Let $\Sigma$ be an arbitrary surface, which may be spacelike or
timelike. Let $n_a$ be the unit normal vector to this surface,
$n_a n^a=\pm 1$, and $n^A_{A'}=n^a e_{aA'}^A$ its spinorial
representation. $n^A_{A'}$ can be considered as an Hermitian
metric for spinors on $\Sigma$, which allow us to introduce an
operation of Hermition conjugation for such spinors
\begin{equation}
\mu_A^\dagger = n_A^{A'} \bar \mu_{A'},
\end{equation}
where bar means complex conjugation. This operation is involutive
$(\mu_A^\dagger)^\dagger = \pm \mu_A$, where '+' stays for a
timelike surface and '-' for a spacelike one.

The operation of Hermition conjugation allows one to define a new
type of connection on the surface $\Sigma$. In four dimensions the
only relevant completely covariant connection is the torsion free
one. We will denote it simply by $\nabla$:
\begin{equation}
\nabla \wedge  e_A^{A'}=de_A^{A'}+A^{-B}_A \wedge
e_B^{A'}-e_A^{B'} \wedge A^{+A'}_{B'}=0.
\end{equation}
Here $A^{-B}_A$ and $A^{+A'}_{B'}$ are anti-self-dual and
self-dual parts of the torsion-free connection which act on
unprimed (left-handed) and primed (right-handed) spinors
respectively. They are related to each other by complex
conjugation:
\begin{equation}
A^{+A'}_{B'}=\bar A^{-A'}_{B'}.
\end{equation}
Along with the torsion-free covariant derivative one can define
define purely anti-self-dual and purely self-dual covariant
derivatives. Let $\mu^{AA'}$ be an arbitrary spinor with one
primed and one unprimed indices. Then we define
\begin{eqnarray}
\nabla^{-}_a \mu^{AA'} &=&\partial_a \mu^{AA'}
+A^{-A}_{aB}\mu^{BA'}
\nonumber \\
\nabla^{+}_a \mu^{AA'} &=&\partial_a \mu^{AA'}
+A^{+A'}_{aB'}\mu^{AB'}. \label{lrcd}
\end{eqnarray}
These ``covariant'' derivatives are not completely covariant. The
first of them restricts the gauge covariance to anti-self dual
transformations and the second to self-dual. However they may give
rise to fully covariant derivatives when projected on the surface.

Below, unless otherwise stated, $\Sigma$ is a spaclike slice of
spacetime, $n_{A'B}$ is the spinorial representation of timelike
unit normal vector and
$e^A_B=e^{AA'}n_{A'B}-\frac{1}{2}e^{CA'}n_{A'C}\delta^A_B$ is the
triad on $\Sigma$.

Let us first show that the torsion of (anti)self-dual connection
defined by (\ref{lrcd}) projected on a timelike surface $S$ is ADM
energy-momentum (generally $S$ is supposed to be taken to infinity
although it doesn't necessarily have to). Let us consider the
second term in the r.h.s of (\ref{actions}) which is the only term
dependent on metric needed for consistency of the action principle
if we don't care about covariance. If we take a spacial slice
$\Sigma$ and make a 3+1 decomposition this action will read
\begin{equation}
S_S=...+\int dt \int e_{tA}^{A'} \wedge A^A_B \wedge e^B_{A'}.
\end{equation}
ADM energy and momentum are coefficients in front of lapse and
shift functions in the above integral, and given that
$e_{tA}^{A'}n^{A'}_A=N$ and $e_{tA}^{A'}e^{A'}_{iA}=N_i$ they are
\begin{eqnarray}
E_{ADM}=\frac{1}{4\pi G}\int_{S \cap \Sigma} e^A_B \wedge A^B_A
\nonumber
\\
(P_{ADM})^A_B=\frac{1}{4\pi G} \int_{S\cap \Sigma}( e^A_C \wedge
A^C_B-\frac{1}{2} e^C_D \wedge A^D_C \delta^A_B), \label{adm}
\end{eqnarray}
where the symmetric pare of spinorial indices $A,B$ labels 3
spacelike directions. The same expressions written in Ashtekar
canonical variables can be found e.g. in \cite{husain}. Now taking
into account that the total connection is torsion-free it is easy
to see that the expressions entering the first and the second
integral in (\ref{adm}) are components of the torsion of the
connection (\ref{lrcd}), $\nabla^+ \wedge e_{A}^{A'}$, projected
on $n^{A'}_A$ and orthogonal to $n^{A'}_A$ respectively.

The above expressions for energy and momenta are simple and have
been proved to be zero in the vacuum, however they are not
covariant and do not form any algebraic structure from the point
of view of boundary theory. Below we consider covariant
expressions  for energy and momenta given by the boundary
constraints (\ref{bconst}).

First, let us notice that the constraints from the first line in
(\ref{bconst}) define quasilocal angular momenta of the bulk
theory.
\begin{equation}
J_{B}^A=\frac{1}{4\pi G}\int_{\Sigma \cap S
}\Big(\frac{1}{\Lambda}F_{B}^A+\sigma^A_{C}\wedge
\sigma^{C}_B\Big) \label{qam}
\end{equation}
Indeed $C^A_B$ in (\ref{bconst}) are boundary terms resulting from
the variation of the Gaussian constraints of the bulk theory,
generating local Lorentz transformations. One can fix a tetrad on
the boundary so that it include gauge condition aligning intrinsic
Lorentz frame with the global basis of boundary spacetime. So
intrinsic Lorentz transformation identified with global ones and
the operator generating them becomes the angular momentum of the
theory.

The rest of constraints (\ref{bconst}) are components of torsion
of purely self-dual connection the triad $\sigma^A_B$ which is the
tetrad $e^{AA'}$ projected on $S$. The connection entering this
torsion can be equally understood as the four-dimensional
torsion-free connection:
\begin{equation}
\nabla^{-} \wedge \sigma^A_B =\nabla \wedge \sigma^A_B.
\label{torss}
\end{equation}
From the fact that the four-dimensional covariant derivative
annihilates the spacetime tetrad $e^{AA'}$ it follows that only
the derivative of the normal vector $n_{A'B}$  contributes to the
torsion (\ref{torss}). This means that the torsion(\ref{torss}) is
related to the extrinsic curvature of $\Sigma$.
\begin{eqnarray}
\nabla \wedge \sigma^A_B &=&\nabla (n_{A'B}) \wedge e^{AA'}=
(dn_{A'B} +A^{-C}_Bn_{A'C}-A^{+C'}_{A'}n_{C'B})n^{A'}_D \wedge
\sigma^{AD} \nonumber \\
 &=&(A^{+A}_C-A^{-A}_C)\wedge \sigma^C_B \label{torss1}
\end{eqnarray}
Here in the last line we introduced a self-dual connection acting
on left-handed spinors
\begin{eqnarray}
A^{+A}_B =n^A_{A'}dn^{A'}_B-n^A_{B'}A^{+B'}_{A'}
n^{A'}_{C}=n^A_{A'}\nabla^+n^{A'}_B \label{asd}
\end{eqnarray}
There are some simple relations between self-dual and
anti-self-dual connections acting on spinors of the same
chirality. First, it follows from (\ref{torss1}) that
\begin{equation}
A^{+A}_B-A^{-A}_C = iK^A_B, \label{excur}
\end{equation}
where $K^A_B$ is the tensor of extrinsic curvature of $\Sigma$. On
the other hand the usual reality condition for Ashtekar variables
means that
\begin{equation}
A^{+A}_B+A^{-A}_C = \Gamma^A_B,
\end{equation}
where $\Gamma^A_B$ is the connection which is torsion-free on
$\Sigma$:
\begin{equation}
\nabla_\Gamma \wedge \sigma^A_B=0.
\end{equation}

Similarly one can introduce an anti-self-dual connection acting on
right-handed spinors
\begin{eqnarray}
A^{-A'}_{B'} = n^{A'}_{A}\nabla^-n^{A}_{B'}.
\end{eqnarray}

Now we should relate $H^A_B$ in (\ref{bconst}) to energy and
momenta. Let $\tau_{\mu A}^B$, $\mu=0,1,2$ be 3 generator of
SL(2,R) group, which is the restriction of SL(2,C) on timelike
slice, $\mu=0$ corresponding to rotation and $\mu=1,2$
corresponding to boosts. By using (\ref{torss1}) and (\ref{excur})
it is easy to show that
\begin{eqnarray}
i \tau_{0 A}^B *(\nabla \wedge \sigma^A_B)\big\vert_{\Sigma \cap
S} = \tau_{0 A}^B *(K^A_C \wedge \sigma^C_B)\big\vert_{\Sigma \cap
S}=det(\sigma\big\vert_{\Sigma \cap S}) K_\alpha^\alpha \\
\nonumber i \tau_{i A}^B *(\nabla \wedge
\sigma^A_B)\big\vert_{\Sigma \cap S} = \tau_{i A}^B*(K^A_C \wedge
\sigma^C_B)\big\vert_{\Sigma \cap S}=det(\sigma\big\vert_{\Sigma
\cap S}) K_t^i, \label{qemd}
\end{eqnarray}
where * denotes hodge dual with respect to the volume form on
$\Sigma \cap S$. In the r.h.s. of the first line of (\ref{qemd})
we recognize the density which when integrated over $\Sigma \cap
S$ give rise to the Brown-York quasilocal energy \cite{brownyork}
and in the second line we find quasilocal momentum. Thus we can
write down the relation between constraints $H^A_B$ from
(\ref{bconst}) and quasilocal energy-momentum as follows
\begin{eqnarray}
E&=&i\tau_{0 A}^B \frac{1}{4\pi G}\int \nabla \wedge \sigma^A_B, \nonumber \\
P_i&=&i\tau_{i A}^B \frac{1}{4\pi G}\int \nabla \wedge \sigma^A_B.
\label{byem}
\end{eqnarray}
The expressions are not simply related to the basic canonical
variables of the bulk theory and they do not vanish in the vacuum.
However they can be related to simple ADM expressions (\ref{adm})
by using a reference spacetime \cite{brownyork}
\begin{eqnarray}
E_{ADM}=E-E_{ref}\nonumber \\
(P_{ADM})^A_B \tau_{i A}^B =P_i-(P_i)_{ref},
\end{eqnarray}
where the subscript $ref$ means 'calculated in a reference
spacetime'. Thus simple form can be restored at the cost of
covariance. The advantage of this form is that quantities with
subscript $ref$ are non-dynamical (they are c-numbers), and we can
calculate bulk commutators of quasilocal quantities by using
simple ADM expressions. More detailed description of Brown-York
energy in Ashtekar variables can be found in the paper of Lau
\cite{lau}.

\section{Einstein's equations as a local conservation law}
It is interesting to notice that the torsion of the connection
(\ref{lrcd}) with the spacetime tetrad
\begin{equation}
T^{A'}_A=\nabla^- \wedge e^{A'}_A
\end{equation}
projection of which on a boundary defines the ADM energy is a
locally conserved quantity in a covariant sense. Indeed the
covariant divergence of $T^{A'}_A$ vanishes due to Einstein's
equations:
\begin{equation}
\nabla \wedge T^{A'}_A = \nabla \wedge \nabla^- \wedge e^{A'}_A=
F^{-B}_A \wedge  e^{A'}_B =0. \label{conservt}
\end{equation}
Here $F^{-B}_A$ is the curvature of the self-dual connection. The
quantities $C^A_B$ from (\ref{bconst}) which define quasilocal
angular momentum when projected on the boundary  are also locally
conserved due to the total connection being torsion-free:
\begin{equation}
\nabla \wedge C^A_B =\nabla \wedge \Big( \frac{1}{\Lambda}F^A_B+
\sigma^A_C \wedge \sigma^C_B \Big)= \nabla \wedge ( e^A_{A'}
\wedge e^{A'}_B )=0.  \label{conservc}
\end{equation}
From (\ref{conservt},\ref{conservc}) it follows that the complete
set of equations of GR is simply equivalent to the condition of
conservation of $T^{A'}_A$ and $C^A_B$. Therefore all the
Einstein's equations can be put in a form of a local conservation
law.

One of the conserved quantities, $T^{A'}_A$, is not a tensor (it
does not transform covariantly with respect to right-handed gauge
transformations). This is like the divergence-freeness  condition
of stress-energy tensor can be reexpressed as a genuine
conservation law for some pseudotensor. In the present situation
we can however rewrite all the equation as a conservation of a
covariant quantities. This can be done on an arbitrary slice of
spacetime. To each such slice one can associate a triad
$\sigma^A_B$ which is the projection of the tetrad $e^A_{A'}$ on
it. The torsion of $\sigma^A_B$, $H^A_B=\nabla \wedge \sigma^A_B$,
is a covariant quantity. In particular, if the slice is timelike
this is a constraint (\ref{bconst}) of the boundary theory
(\ref{actionsc}). The covariant divergence of $H^A_B$ is
equivalent to a subset of Einstein's equations:
\begin{equation}
\nabla \wedge H^A_B= \nabla \wedge \nabla \wedge \sigma^A_B =
F^A_C \wedge \sigma^C_B=0 \label{conservh}
\end{equation}
These equations are not necessarily defined on a single slice. One
can consider one parameter family of slices or foliation of the
whole spacetime to define it everywhere. One foliation is however
not enough to recover the whole set of Einstein's equations
(equations with components normal to the foliation are still
missing). At least two foliations which are different everywhere
are required.

The equations (\ref{conservt},\ref{conservc}) pulled back on a
spacial slice $\Sigma$ form the complete set of constraints of GR.
Formally all the constraints are now Gaussian -- they say that the
electric field $C^A_B$ and some other field $T^{A'}_A$ are
divergence-free. Quantum-mechanically this form of constraints is
very difficult to treat because of complicated dependance of
$H^A_B$ and even $T^{A'}_A$ on the basic canonical variables.
However it provides us with some intuition about what generic
solutions of the bulk theory must look like. If we introduce a
perturbation on the boundary which has certain energy and momentum
it has to be continued into the bulk as a certain flux of
energy-momentum. Constraint equations simply mean that the lines
of such fluxes must be continuous.

\section{Discussion: Towards the definition of excitations around the vacuum}
In this section we are making a proposal on how to define
excitations around the vacuum which will be developed in
subsequent papers.

Any physical excitation must have nonzero energy. As the energy is
defined on a boundary of spacetime, any excitation must be seen
from the boundary. Energy together with other boundary
observables, such as momentum and angular momentum, provide
exactly enough information to specify any excitation. Therefore to
introduce an excitation in the vacuum one just has to change
somehow the value of the boundary observables.

A natural (and in fact the simplest) way to give a nonzero value
to the boundary observables is to put up a topological puncture on
the boundary. Because the symmetry of the boundary theory is
SO(3,1), containing rotations and boosts parallel to the boundary,
and translations tangential to it, to each puncture one can
associate a representation of SO(3,1) group (a charge of SO(3,1)
gauge field) labeled by casimirs such as mass and spin, and a
vector in this representation. This vector specifies  the momentum
and the polarization of a particle moving along the boundary.

The constraints of general relativity must then continue the
excitation introduced on the boundary into the bulk. Because
puncture on the boundary is an SO(3,1) charge it must induce a
flux of SO(3,1) gauge field. As it was discussed in section 5 the
constraints simply mean that the lines of such fluxes must be
continuous.

The continuous lines of fluxes of a gauge field may be expressed
as Wilson lines of this field. Now the gauge group is SO(3,1) and
the Wilson line is
\begin{equation}
W(a) =P e^{\int a_a dx^a}, \label{ncloop}
\end{equation}
where the connection $a$ is that from (\ref{so31c}).

The loops (\ref{ncloop}) are very difficult to treat
quantum-mechanically, because the connection $a$ entring them is
non-commutative with respect to the bulk simplectic form and the
expression for a commutator between different components of this
connection is very involved. It is therefore difficult to say if
the loops (\ref{ncloop}) indeed satisfy all the constraints of
general relativity or not. Yet (although in a different context) a
formalism treating non-commutative loops has been developed
\cite{alexand} and some definite physical predictions have been
extracted from it.

The bulk equations of motion can in turn put some constraints on
the boundary observables, in particular they can relate masses and
spins of particles propagating along the boundary. A glimpse of
such a constraint can be found in the work of Thiemann
\cite{thiemannadm} on ADM energy of spin-network states. The loops
with commutative connection considered there are special cases of
loops (\ref{ncloop}) where the representation is labeled by only
one casimir which is the spin $j$. In \cite{thiemannadm} it was
shown that in a simple case in which the ADM energy can be
diagonalized its enigenvalues scale like
\begin{equation}
E_{ADM} \sim \sqrt[4]{j(j+1)}. \label{evj}
\end{equation}
Similar expression for quasilocal energy was obtained in
\cite{major} for a completely different regularization. So the
relation (\ref{evj}) seems to be generic.
 When momentum is zero the energy
is proportional to the mass of the particle and as a consequence
for large spins we have the following relation
\begin{equation}
m^2 \sim  j.
\end{equation}
This is Regge type of relation for string oscillation modes. This
may indicate that some sort of string theory may arise here. Of
course this doesn't have to be any kind of critical strings so far
known.

The problem with (\ref{evj}) is that the energy spectrum is
discrete and therefore cannot describe propagating particles (they
must have continuous kinetic energy). This can be solved by
including the more general type of loops (\ref{ncloop}) for
SO(3,1) group for which the spectrum must be continuous. The last
statement is somehow supported by some of the results of
\cite{alexand}.

All the above will be developed in subsequent papers.

\section*{Acknowlegements}

I would like to thank Lee Smolin and Laurent Freidel for extensive
conversations and helpful comments on the draft, and  Hideo Kodama
and Karim Noui for useful discussions on some aspects of this
work.


\begin{thebibliography}{99}
\bibitem{reviewrt}
 C.~Rovelli, \emph{Loop Quantum Gravity.} Living Reviews in Relativity,
        1998-1, http://www.livingreviews.org/ and e-print: gr-qc/9710008.

T.~Thiemann,  Lectures on Loop Quantum Gravity,  gr-qc/0210094


\bibitem{Thiemann96} T.~Thiemann, \emph{Anomaly-free formulation of non-perturbative,
    four-dimensional Lorentzian quantum gravity.} Phys. Lett. {\bf B380} (1996) 257,
        e-print: gr-qc/9606088.

 T.~Thiemann, \emph{Quantum Spin Dynamics (QSD)}
        Class. Quant. Grav. {\bf 15} (1998) 839, e-print: gr-qc/9606089.


\bibitem{Smolin96}
L.~Smolin, The classical limit and the form of the hamiltonian
constraint in nonperturbative quantum gravity, gr-qc/9609034



\bibitem{LewandowskiMarolf}\comment{online}
\comment{author}Lewandowski, J., and Marolf, D.,
\comment{onlinetitle}``Loop
  constraints: A habitat and their algebra'', (\comment{onlineyear}1997),

\comment{author}Gambini, R., Lewandowski, J., Marolf, D., and
Pullin, J.,
  \comment{onlinetitle}``On the consistency os the constraint algebra in spin
  network quantum gravity'', (\comment{onlinemonth}September,
  \comment{onlineyear}1997),


\bibitem{Thiemann99}
T. Thiemann,  Gauge Field Theory Coherent States (GCS) : I.
General Properties,  hep-th/0005233,

 T. Thiemann, O. Winkler,  Gauge Field Theory Coherent States (GCS) : III. Ehrenfest
 Theorems,  hep-th/0005234

 T. Thiemann, O. Winkler,  Gauge Field Theory Coherent States (GCS) : IV. Infinite Tensor Product and Thermodynamical
 Limit,  hep-th/0005235

 T. Thiemann, O. Winkler, Gauge Field Theory Coherent States (GCS) : II. Peakedness
 Properties, hep-th/0005237.


\bibitem{Varadarajan2000}
M.~Varadarajan, \emph{ Fock representations from U(1) holonomy
algebras}, Phys. Rev {\bf D61}, 104001 (2000).

M.~Varadarajan, \emph{ Photons from quantized electric flux
representations}, gr-qc/0104051

\bibitem{Kodama90}
H.~Kodama, Prog. Theor. Phys 80, 1024 (1988); Phys. Rev.
D42(1990)2548.


\bibitem{Ashtekar}
A. Ashtekar, New Variables for Classical and Quantum Gravity,
Phys. Rev. Lett. 51(18), 2244-2247, (1986)

\bibitem{Smolin02}
L.~Smolin, Quantum gravity with a positive cosmological constant,
hep-th/0209079

\bibitem{HooftDJ84} S. Deser, R. Jackew, and G. t'Hooft,
Ann. Phys. 152 (1984) 220.

\bibitem{JacobsonSmolin88} T.~Jacobson and L.~Smolin, \emph{Nonperturbative quantum geometries.}
    Nucl. Phys. {\bf B299} (1988) 295.

\bibitem{RovelliSmolin88} C.~Rovelli and L.~Smolin, \emph{Knot theory and quantum
    gravity.} Phys. Rev. Lett. 61 (1988) 1155.

 C.~Rovelli and L.~Smolin, \emph{Loop representation of quantum
    general relativity.} Nucl. Phys. {\bf B331} (1990) 80.

\bibitem{Rovelli98}
C.~Rovelli, \emph{The projector on physical states in loop quantum
gravity}, gr-qc/9806121

\bibitem{Smolin95}
L. Smolin, Linking Topological Quantum Field Theory and
Nonperturbative Quantum Gravity,  J.Math.Phys. 36 (1995)
6417-6455, gr-qc/9505028

\bibitem{Smolin98}
L. Smolin, {\it A holographic formulation of quantum general
relativity}, hep-th/9808191,  Phys.Rev. D61 (2000) 084007; Yi
Ling, Lee Smolin, {\it Holographic Formulation of Quantum
Supergravity},
 hep-th/0009018,  Phys.Rev. D63 (2001) 064010


\bibitem{Krasnov96}
K. Krasnov,  On Quantum Statistical Mechanics of a Schwarzschild
Black Hole,  Gen.Rel.Grav. 30 (1998) 53-68, gr-qc/9605047

\bibitem{mmons} S. Macdowell, F. Mansouri, Phys.Rev.Lett. 38:739,
1977, Erratum- ibid.38:1376,1977


\bibitem{witten89}
 E. Witten,
Nucl. Phys. B331 (1998) 46.

\bibitem{husain}
 V. Husain, S. Major, Gravity and BF theory defined in bounded
 regions, Nucl.Phys. B500 (1997) 381-401, gr-qc/9703043

\bibitem{brownyork}
J. D. Brown, J. W. York, Quasilocal Energy and Conserved Charges
Derived from the Gravitational Action, gr-qc/9209012



\bibitem{lau}
 S. R. Lau, New variables, the gravitational action, and boosted quasilocal
 stress-energy-momentum, Class.Quant.Grav. 13 (1996) 1509-1540,  gr-qc/9504026





\bibitem{dewitt}
B. DeWitt,  QUANTUM THEORY OF GRAVITY. 1. THE CANONICAL THEORY.,
Phys.Rev.160:1113-1148,(1967)


\bibitem{alexand}
 S. Alexandrov, D. Vassilevich, Area spectrum in Lorentz covariant loop
 gravity, Phys.Rev. D64 (2001) 044023, gr-qc/0103105,

 S. Alexandrov, E. R. Livine, SU(2) Loop Quantum Gravity seen from Covariant
 Theory,  Phys.Rev. D67 (2003) 044009, gr-qc/0209105

\bibitem{thiemannadm}
T. Thiemann, QSD VI : Quantum Poincar\'e Algebra and a Quantum
Positivity of Energy Theorem for Canonical Quantum Gravity,
Class.Quant.Grav. 15 (1998) 1463-1485,  gr-qc/9705020




\bibitem{major}
S.Major,  Quasilocal Energy for Spin-net Gravity,
Class.Quant.Grav. 17 (2000) 1467-1487,  gr-qc/9906052






















\end{thebibliography}
\end{document}